\documentclass[11pt,draftclsnofoot,journal,onecolumn]{IEEEtran}

\IEEEoverridecommandlockouts

\usepackage{graphicx}
\usepackage{cite}
\usepackage{url}
\usepackage[cmex10]{amsmath}
\usepackage{amssymb}
\usepackage{algorithm}
\usepackage[noend]{algorithmic}
\usepackage{cases}
\usepackage[caption=false,font=footnotesize]{subfig}

\DeclareMathOperator*{\argmax}{arg\,max}
\DeclareMathOperator{\pw}{P}
\DeclareMathOperator{\edge}{edge}

\DeclareMathOperator{\pollcluster}{PollCluster}
\DeclareMathOperator{\average}{average}
\DeclareMathOperator{\nex}{next}
\DeclareMathOperator{\adj}{adj}
\DeclareMathOperator{\head}{head}
\DeclareMathOperator{\E}{\mathbb{E}}

\begin{document}

\title{Weighted Centroid Localization Algorithm: Theoretical Analysis and Distributed Implementation}
\author{Jun~Wang, Paulo~Urriza, Yuxing~Han, and Danijela~\v{C}abri{\'c}%
\thanks{The authors are with the Department of Electrical Engineering, University of California, Los Angeles, CA, 90095, USA (email: \{eejwang, pmurriza, ericahan, danijela\}@ee.ucla.edu).}
\thanks{Part of this work has been accepted to the proceedings of the Asilomar Conference on Signals, Systems, and Computers, Nov. 7--10, 2010, Pacific Grove, CA, USA \cite{Wang2010}.}
\thanks{This work has been submitted to the IEEE for possible publication. Copyright may be transferred without notice, after which this version may no longer be accessible.}}

\maketitle


\begin{abstract}
Information about primary transmitter location is crucial in enabling several key capabilities in cognitive radio networks, including improved spatio-temporal sensing, intelligent location-aware routing, as well as aiding spectrum policy enforcement. Compared to other proposed non-interactive localization algorithms, the weighted centroid localization (WCL) scheme uses only the received signal strength information, which makes it simple to implement and robust to variations in the propagation environment.  In this paper we present the first theoretical framework for WCL performance analysis in terms of its localization error distribution parameterized by node density, node placement, shadowing variance, correlation distance and inaccuracy of sensor node positioning.  Using this analysis, we quantify the robustness of WCL to various physical conditions and provide design guidelines, such as node placement and spacing, for the practical deployment of WCL. We also propose a power-efficient method for implementing WCL through a distributed cluster-based algorithm, that achieves comparable accuracy with its centralized counterpart.
\end{abstract}

\IEEEpeerreviewmaketitle


\section{Introduction}
\label{sec:Introduction}

Cognitive Radio (CR) is a promising approach to efficiently utilize the scarce RF spectrum resource. In this paradigm, knowledge about spectrum occupancy in time, frequency, and space that is both accurate and timely is crucial in allowing CR networks to opportunistically use the spectrum, and avoid interference to a primary user (PU). In particular, information about PU location will enable several key capabilities in CR networks including improved spatio-temporal sensing, intelligent location-aware routing, as well as aiding spectrum policy enforcement.

The localization problem in CR networks is in general different from localization in other applications such as Wireless Sensor Networks (WSN) and Global Positioning System (GPS), in which the target to be localized cooperates with the localization devices. In contrast, a PU does not communicate directly with the CRs during the localization process. This scenario is referred to as non-cooperative localization and does not allow the use of conventional techniques such as time-of-arrival (TOA) and time-delay-of-arrival (TDOA) ranging.

\subsection{Related Work}

Several techniques have been proposed in the literature for estimating the position of active PUs in a CR network. These schemes include both range-based \cite{Langendoen2003,Xiao2007} and range-free \cite{Bulusu2000,He2003,Liu2004,Shang2003,Ma2010,Chen2010a,Chen2010b} algorithms. Range-free positioning schemes, such as centroid localization \cite{He2003}, have attracted a lot of interest because of their simplicity and robustness to changes in wireless propagation properties such as path loss. This characteristic makes them suitable candidates for systems requiring coarse-grained, but reliable and cost-effective techniques. On the other hand, range-based techniques such as multi-lateration \cite{Langendoen2003}, offer better estimates, but require accurate information about the target's distance that can only be achieved either through cooperation with the target or a very precise knowledge of the path loss model.

In this paper, we focus on analyzing the performance of a particular, low complexity, coarse-grained localization algorithm, referred to as Weighted Centroid Localization (WCL) \cite{Blumenthal2007}. In this technique, PU location is approximated as the weighted average of all secondary user positions within its transmission range. Strictly speaking, WCL is not an entirely range-free technique because it requires additional information aside from simple connectivity, namely Received Signal Strength (RSS) measurements. WCL and WCL-like methods are suitable for PU localization because of their low computational complexity. They also do not require cooperation from the PU and only use the readily available RSS information. In addition, WCL always converges to a solution, as opposed to some range-based techniques, and do not require estimation of the path loss exponent beforehand.

Recent papers on WCL have proposed improvements on the basic algorithm to enhance its performance for specific scenarios \cite{Laurendeau2010,Liu2009,Behnke2009,Chen2010}. All previous WCL studies reported in the literature have been analyzed either via simulation or field experiments \cite{Blumenthal2007,Salomon2005}. To the best of our knowledge, there has been no analysis of the probability distribution of the localization error for even the simplest version of WCL in the presence of shadowing in both WSN and CR literature.

\subsection{Contributions}

We present the first theoretical analysis of the error distribution of WCL in this paper. The presented analytical framework models varying levels of shadowing, including both independent and correlated shadowing environments, and also takes inaccuracy of sensor node positioning into consideration. Using this analysis in conjunction with numerical simulations we observe and quantify the robustness of WCL to variable channel conditions and node placements. We also investigate variable node participation as a simple modification that can improve WCL performance.

In the latter part of this paper we propose and investigate a distributed cluster-based implementation of WCL in the context of CR networks. Previous work primarily assumes a centralized fusion node aggregating data from other nodes. In the context of WSN where WCL was first utilized \cite{Blumenthal2007}, the primary motivation was node self-localization where a node applies WCL using measurements from beacon nodes in close proximity to it. This approach naturally dictates a centralized system where the node to be localized acts as the fusion center and only requires minimal interaction with the beacons. In the case of PU localization in CR networks, however, where the primary motivation is estimating the position of a non-cooperative node, there no longer exists any particular node that has all the requisite information to perform WCL. In this case a fusion node is arbitrarily chosen. This approach does not scale very well in terms of communication overhead and transmit power, which limits its application to low node densities or small network size.

There is therefore a need to create distributed implementations of WCL for this technique to be practically applied in CR localization. A standard approach for distributing control in sensor networks is clustering. Several algorithms exist in the literature for creating either disjoint \cite{Abbasi2007,Zhang2002} or overlapping \cite{Youssef2009} clusters. In \cite{Zhou2009}, a distributed technique based on the concept of mean-shift \cite{Cheng1995} was proposed. However, the technique requires repeatedly forming a virtual cluster until it converges to the target location. This introduces additional communication overhead and also requires several iterations to converge. We present an implementation which saves total transmit power needed for localization process communication, and investigate its accuracy as well as its effect on per node computational complexity and total transmit power consumption.

The rest of the paper is organized as follows. The system model is introduced in Section \ref{sec:SystemModel}. The theoretical analysis of centralized WCL accuracy is presented in Section \ref{sec:Analysis}, while a practical and distributed implementation of WCL is proposed and analyzed in Section \ref{sec:Distributed}. Numerical results for various scenarios used to evaluate performance of WCL algorithms are presented in Section \ref{sec:Evaluation}. Finally, the paper is concluded in Section \ref{sec:Conclusion}.


\section{System Model}
\label{sec:SystemModel}
This section presents the general framework for our analysis. We first review the assumptions about the system setting and propagation model, and then present the basic WCL algorithm.

We adopt the PU-centric model \cite{Liu2009} in which the PU transmitter is assumed to be at the center area of a circle. The rationale behind this choice is that in a practical scenario, sensor nodes that obtain enough received power (i.e. above sensitivity threshold) approximately form a circle around the primary transmitter. Thus, sensors within this circle are used to jointly estimate the position of the PU. An implicit assumption in this model is that the PU is surrounded by enough nodes so that edge effects can be disregarded.

Assume $N$ sensor nodes in a circle of fixed radius $R$. The number of sensors within the circle is jointly affected by several factors, such as transmit power of the PU (which effectively determines $R$), node density, node selection methodology, etc. $N$ is a design parameter for WCL deployment, the choice of which will have significant influence on the localization accuracy, as will be shown in our analysis.

The PU located at the center has a coordinate defined as $\mathbf{L}_p\triangleq [x_p,y_p]^T=[0, 0]^T$. For the $i$th sensor node, we define its 2-dimensional location as $\overline{\mathbf{L}}_i = \left[  \overline{x}_i, \overline{y}_i \right]^T$ where $\overline{x}_i^2 + \overline{y}_i^2 \leq R^2$. Sensors are placed independently in both dimensions. We assume sensors obtain information about their own locations, which is modeled as imperfect. The measured location of the $i$th sensor is distributed as $\mathbf{L}_i \sim \mathcal{N} (\overline{\mathbf{L}}_i, \sigma_l^2 \textbf{I}_2)$, where $\sigma_l$ defines standard deviation of the error in sensor node self-localization. We further assume the self-localization errors are independent among sensors and are independent from received powers.

We adopt the following channel model for received signal strength used in WCL. The received power of the $i$th node from the primary transmitter, $P_i$, is given by
\begin{equation}\label{received power}
   P_i = P_0-10\gamma\log\left(\frac{\| \overline{\mathbf{L}}_i-\mathbf{L}_p \|}{d_0}\right)+s_i \; \text{dB},
\end{equation}
in which the first two terms characterize path loss and the last
term describes shadowing effect. The path loss is characterized by
three constants: reference power $P_0$, reference distance $d_0$ and
path loss exponent $\gamma$. Denote
collection of shadowing variables as $\mathbf{s}\triangleq \left[s_1,s_2,\ldots,s_N\right]$, then its distribution is characterized by $\mathbf{s} \sim \mathcal{N} (\bf{0},
\mathbf{\Omega}_\mathbf{s})$. We consider the following shadowing cases
\begin{subnumcases}{\mathbf{\Omega}_\mathbf{s}=}
    \sigma_s^2 \bf{I}_N, & i.i.d case,\\
    \{\mathbf{\Omega}_\mathbf{s}\}_{ij} = \sigma_s^2 e^{- \| \overline{\mathbf{L}}_i-\overline{\mathbf{L}}_j \| /X_c}, & correlated case,
\label{shadowing}
\end{subnumcases}
where $X_c$ is the correlation distance \cite{Gudmundson1991} within which the shadowing effects of nodes are correlated, and $\bf{I}_N$ is the identity matrix with dimension $N$.

The theoretical analysis presented in Section \ref{sec:Analysis} is based on the centralized architecture, which will also give us an upper bound on the performance. In each localization period, all sensors involved in the WCL will send their RSS measurements to the fusion center, where an estimate will be made about the coordinate of the primary transmitter. The considered algorithm is a modification of relative-span WCL proposed in \cite{Laurendeau2010}. In this scheme, the weights are guaranteed to be non-negative in dB scale. The 2-dimensional estimated location of PU is formed as
\begin{equation}\label{est location}
  \hat{\mathbf{{L}}}_p = \frac{\sum_{i=1}^{N} w_i \mathbf{L}_i}{\sum_{i=1}^{N} w_i}   = \frac{\sum_{i=1}^{N} \left(P_i - P_{\min}\right) \mathbf{L}_i}{\sum_{i=1}^{N}(P_i -
  P_{\min})}.
\end{equation}
The weight factor of node $i$ is defined as $w_i = (P_i - P_{\min})/\Delta P$, $i=1,2,\ldots,N,$ where $P_{\min}$ is a constant calculated from average received power of the node on the border with some margin, to guarantee the probability that receive power of all nodes in the area below $P_{\min}$ is sufficiently small, say 1\%. $P_{\max}$ is the maximum received power among nodes, and $\Delta P\triangleq P_{\max}-P_{\min}$ is the span of received
power. The localization error is then given by $\mathbf{e}_L \triangleq \hat{\mathbf{L}}_p - \mathbf{L}_p = \hat{\mathbf{L}}_p = \left[\hat{x}_p, \hat{y}_p\right]^T$, where $\hat{x}_p$ and $\hat{y}_p$ are the 1-dimensional errors along the x and y-axis respectively. Finally, the performance of the WCL scheme is evaluated by distance error, which is given by
\begin{equation}\label{error-dist}
  e_{L} \triangleq \sqrt{\hat{x}_p^2 + \hat{y}_p^2} = {\| \mathbf{e}_L \|}_2.
\end{equation}


\section{Theoretical Analysis of WCL Error Distribution}
\label{sec:Analysis}
This section presents the analysis of WCL error using probability theory, ultimately leading to analytical expressions for the error distribution, first of WCL applied to 1-dimensional coordinates, and finally the 2-dimensional WCL error given by (\ref{error-dist}).

\subsection{1-dimensional Location Estimation Error}
\label{subsec:1dError}

\subsubsection{I.I.D Shadowing}
\label{subsubsec:IIDShadowing}

First we derive the 1-dimensional localization error for the i.i.d. shadowing case. We define a constant $\mu_i$ for
each node $i$ as
\begin{equation}\label{constant}
   \mu_i \triangleq P_0-10\gamma\log\left(\frac{\| \overline{\mathbf{L}}_i-\mathbf{L}_p
   \|}{d_0}\right)-P_{\min},
\end{equation}
One can verify that $P_i -
P_{\min}=\mu_i + s_i \sim \mathcal{N}\left(\mu_i, \sigma_s^{2}\right)$. The location estimation in one dimension, say, x-axis, is naturally given by
\begin{equation}\label{est x}
   \hat{x}_p = \frac{\sum_{i=1}^{N} \left(P_i - P_{\min}\right) x_i}{\sum_{i=1}^{N}\left(P_i -
   P_{\min}\right)} \triangleq \frac{a}{b}.
\end{equation}
Based on our i.i.d assumption among all $P_i$'s, one can verify that
\begin{equation}\label{distribution b}
   b \sim \mathcal{N}\left(\sum_{i=1}^{N} \mu_i,
   N\sigma_s^{2}\right) \triangleq \mathcal{N}\left(m_b, \sigma_b^2\right).
\end{equation}
For the statistical property of $a$, since $x_i$'s are also i.i.d and are independent of $P_i$'s, the mean of $a$ is given by $m_a = \sum_{i=1}^{N} \mu_i \overline{x}_i$. The variance of the $i$th term of $a$ is given by
\begin{IEEEeqnarray}{rCl}
\label{variance a}
    \text{Var}[(P_i - P_{\min}) x_i] &=& \text{E} [(P_i - P_{\min})^2] \text{E} [x_i^2] - \text{E} [P_i - P_{\min}]^2 \text{E} [x_i]^2 \IEEEnonumber \\
    &=& \sigma_l^2 \mu_i^2 + \sigma_s^2 \overline{x}_i^2 + \sigma_l^2 \sigma_s^2.
\end{IEEEeqnarray}
Thus the variance of $a$ is derived as $\sigma_a^2 = N \sigma_l^2 \sigma_s^2 + \sum_{i=1}^{N} (\sigma_l^2 \mu_i^2 + \sigma_s^2 \overline{x}_i^2)$. From central limit theorem, we approximate $a$ as a single Gaussian variable, i.e. $a \sim \mathcal{N} (m_a, \sigma_a^2)$, which is verified by numerical simulations that are not presented here due to space limit.

Since $a$ and $b$ are correlated due to summation over the same set of variables, i.e. $P_i$'s, $\hat{x}_p$ is a ratio of two correlated Gaussian variables. In order to obtain statistical properties of $\hat{x}_p$, we first calculate correlation coefficient $\rho_{ab}$ of $a$ and $b$, which is given by
\begin{equation}
   \rho_{ab}
      = \E \left[(a-m_a)(b-m_b)\right]/(\sigma_a \sigma_b)
      = \E \left[ \sum_{i=1}^{N} \sum_{j=1}^{N} x_i s_i s_j \right]/(\sigma_a \sigma_b).
      \label{correlation-coeff1}
\end{equation}
Note that
\begin{subnumcases}{\E \left[x_i s_i s_j\right]  =}
    \overline{x}_i \sigma_s^2, & $i=j$,\\
    0, & $i \neq j$,
\label{correlation}
\end{subnumcases}
thus the correlation coefficient is obtained as
\begin{equation}
    \rho_{ab} = \frac {\sigma_s^2 \sum_{i=1}^{N} \overline{x}_i} {\sigma_a \sigma_b}
              = \frac {\sigma_s {\|  \overline{\mathbf{x}} \|}_1}{\sqrt{N^2 \sigma_l^2 \sigma_s^2 + N \sum_{i=1}^{N} (\sigma_l^2 \mu_i^2 + \sigma_s^2 \overline{x}_i^2)}},
\label{correlation coefficient iid}
\end{equation}
where $\overline{\mathbf{x}} \triangleq [\overline{x}_1, \ \ldots \,\overline{x}_N]^T$, and ${\| \mathbf{x} \|}_l$ is the $l$th norm of the vector.

From the previous analysis we know that
$\hat{x}_p$ is the ratio of two dependent Gaussian random variables
with non-zero means for which the closed-form pdf is available in
\cite[Eqn 7.14]{Simon2006}. Based on the pdf of
$\hat{x}_p$, we calculate its mean $m_{\hat{x}_p}$ and variance
$\sigma_{\hat{x}_p}^2$ by numerical integration. Alternatively, we also calculate $m_{\hat{x}_p}$ and
$\sigma_{\hat{x}_p}^2$ through approximation results of \cite{Hayya1975},
which assume $\hat{x}_p$ follows normal distribution. The approximation results are
given by
\begin{IEEEeqnarray}{rCl}
  m_{\hat{x}_p} &\simeq& \left(m_a/m_b\right) + \sigma_b^2 m_a/m_b^3 - \rho_{ab}
  \sigma_a  \sigma_b /m_b^2 \IEEEnonumber \\
  \sigma_{\hat{x}_p}^2 &\simeq& \sigma_b^2 m_a^2/m_b^4 + \sigma_a^2/m_b^2 -
  2 \rho_{ab} \sigma_a  \sigma_b m_a /m_b^3.
  \label{approx}
\end{IEEEeqnarray}
The Gaussian approximation (\ref{approx}) is verified via numerical simulations. Specifically, the mean and variance calculated from integration of the closed-form pdf are compared with the approximated mean and variance, for different node numbers. Our results show that for $N\geq30$, the approximation error for the means is essentially zero and less than 3\% for the variances.

Up to this stage, we are able to calculate the mean and variance of
the estimation error in the x-axis as $m_{e_x} = m_{\hat{x}_p}-x_p = m_{\hat{x}_p}$, $\sigma_{e_x}^2 = \sigma_{\hat{x}_p}^2$. The same analysis applies for calculating the mean and variance of the estimation error along the y-axis, namely, $m_{e_{y}}$ and
$\sigma_{e_{y}}^2$. Note that the estimation errors in x-axis and y-axis are correlated, therefore the statistical properties of the 2-dimensional error, $e_{L}$, cannot be obtained directly.

\subsubsection{Correlated Shadowing}
\label{subsubsec:CorrelatedShadowing}

In this case, since $P_i$'s in (\ref{est x}) are correlated, thus the distribution of $\hat{x}_p$ needs to be evaluated again. Since $b$ in (\ref{est x}) is the linear combination of
correlated Gaussian variables, it is still Gaussian distributed \cite{Proakis2001}, i.e. $b \sim \mathcal{N}\left(m_b, \sigma_b^2\right)$. Thus, $m_b=\sum_{i=1}^{N} \mu_i$, and $\sigma_{b}^2$ is given by $\sigma_{b}^2 = \sum_{i=1}^{N} \sum_{j=1}^{N} \mathrm{R}_{ij}$, where $\mathrm{R}_{ij}$ is the $ij$th element of the covariance matrix of $\left[\left(P_1-P_{min}\right), \ \ldots \ , \left(P_{N}-P_{min}\right)\right]^{T}$, defined as
\begin{equation}
   \mathrm{R}_{ij} = \E \left[\left(P_i-P_{min}-\mu_i\right)\left(P_j-P_{min}-\mu_j\right)\right]
                 = \E [s_i s_j] = \sigma_s^2 \lambda_{ij},
\label{covariance_b}
\end{equation}
where $\lambda_{ij} \triangleq e^{- \| \overline{\mathbf{L}}_i-\overline{\mathbf{L}}_j \| /X_c}$.

For the distribution of $a$, it follows directly that
$m_a=\sum_{i=1}^{N} \overline{x}_i \mu_i$. Its variance is given
by $\sigma_{a}^2 = \sum_{i=1}^{N} \sum_{j=1}^{N} \mathrm{R}'_{ij}$, where $\mathrm{R}'_{ij}$ is the $ij$th element of the covariance matrix of $[(P_1-P_{min})x_1, \ \ldots \ , (P_{N}-P_{min})x_N]^{T}$,
which can be represented by
\begin{equation}
   \mathrm{R}'_{ij} = \E \{[(P_i-P_{min})x_i-x_i \mu_i][(P_j-P_{min})x_j- x_j \mu_j]\} \IEEEnonumber \\
                  = \E [x_i s_i x_j s_j],
\label{covariance_a}
\end{equation}
where $\E [x_i s_i x_j s_j]$ is given by
\begin{subnumcases}{  \text{E} \left[x_i x_j s_i s_j\right]  =}
    \sigma_s^2 (\overline{x}_i^2 + \sigma_l^2) , & $i=j$,\\
    \overline{x}_i \overline{x}_j \sigma_s^2 \lambda_{ij}, & $i \neq j$.
\label{correlation}
\end{subnumcases}
Based on central limit theorem, we approximate $a$ as a Gaussian random variable, i.e. $a \sim \mathcal{N}(m_a, \sigma_a^2)$. The approximation is validated via simulations, and it applies for $N \geq 20$.

The correlation coefficient $\rho_{ab}$ is given by
\begin{equation}
   \rho_{ab} = \E \left[(a-m_a)(b-m_b)\right]/(\sigma_a \sigma_b)
                               = (\sigma_s^2 \sum_{i=1}^{N}
                                   \sum_{j=1}^{N} \overline{x}_i \lambda_{ij}) / (\sigma_a \sigma_b).
\label{correlation coeff}
\end{equation}
Substituting the expressions for $\sigma_a$ and $\sigma_b$ into (\ref{correlation coeff}), after some simple manipulation,
\begin{equation}\label{correlation_coeff1}
   \rho_{ab}=\frac {\mathbf{1}^{T} \mathbf{\Lambda} \overline{\mathbf{x}}} {\sqrt{(\overline{\mathbf{x}}^T \mathbf{\Lambda} \overline{\mathbf{x}} + N \sigma_s^2 \sigma_l^2) \mathbf{1}^{T} \mathbf{\Lambda}
   \mathbf{1}}},
\end{equation}
where $\{\mathbf{\Lambda}\}_{ij}=\lambda_{ij}$ and $\mathbf{1}$ is a $N \times 1$ vector of all $1$'s.

Since the location estimate $\hat{x}_p$ is formed by $\hat{x}_p=a/b$, we use the closed-form pdf given by \cite[Eqn 7.14]{Simon2006} to calculate $m_{\hat{x}_p}$ and $\sigma_{\hat{x}_p}^2$ by numerical integration. Gaussian approximation can also be used similarly to i.i.d shadowing case in Sec. \ref{subsubsec:IIDShadowing}.

\subsection{2-dimensional Location Estimation Error}
\label{subsec:2dError}

In our previous analysis, we obtained the statistical
properties of 1-dimensional location estimation error for both
i.i.d. and correlated shadowing case. From Gaussian approximation of 1-dimensional localization error (\ref{approx}), we know that $\mathbf{e}_L\sim \mathcal{N}(\bar{\mathbf{e}}_L,
   \mathbf{\Omega}_L)$, where $\bar{\mathbf{e}}_L \triangleq [m_{\hat{x}_p}, m_{\hat{y}_p}]^T$,
\begin{equation}
  \mathbf{\Omega}_L \triangleq \left [ \begin{array}{cc}
             \sigma_{\hat{x}_p}^2 & \rho_{\hat{x}_p \hat{y}_p} \sigma_{\hat{x}_p} \sigma_{\hat{y}_p}\\
             \rho_{\hat{x}_p \hat{y}_p} \sigma_{\hat{x}_p} \sigma_{\hat{y}_p} & \sigma_{\hat{y}_p}^2\\
             \end{array}\right],
\end{equation}
and $\rho_{\hat{x}_p \hat{y}_p}$ is the correlation coefficient between
1-dimensional errors, the calculation of which is presented in the Appendix.

The calculation of localization error $e_{L}$ involves computing 2-norm of $\mathbf{e}_L$. Since the two variables in $\mathbf{e}_L$ are dependent Gaussian r.v.'s with different variances, we perform the following transformations. First we form two independent Gaussian variables from
$\mathbf{e}_L$, by the standard technique of decorrelation \cite{Proakis2001}, $\mathbf{e}_L' = \mathbf{Q} \mathbf{e}_L \triangleq \left[\hat{x}_p', \hat{y}_p'\right]^T$, where $\mathbf{Q}$ is the orthogonal eigenvalue decomposition matrix of
$\mathbf{\Omega}_L$, which satisfies $\mathbf{Q} \mathbf{\Omega}_L \mathbf{Q}^{T} = \mathbf{D}$,
where $\mathbf{D}$ is a diagonal matrix. Therefore, one can
verify that \cite[Eqn 2.1-159]{Proakis2001} $\mathbf{e}_L' \sim \mathcal{N} \left(\mathbf{Q} \bar{\mathbf{e}}_L,
  \mathbf{D}\right)$. Then by the orthogonal-invariant property of 2-norm \cite{Proakis2001}, we know $e_L = {\| \mathbf{e}_L \|}_2 = \sqrt{\mathbf{e}_L^{T} \mathbf{Q}^{T} \mathbf{Q}
    \mathbf{e}_L} = {\| \mathbf{e}_L' \|}_2$.

In order to obtain the pdf of ${\| \mathbf{e}_L' \|}_2$, let us first consider pdf of ${\| \mathbf{e}_L' \|}_2^2 ={\hat{x}_p}^{'2} + {\hat{y}_p}^{'2}$. ${\hat{x}_p}^{'2}$ and ${\hat{y}_p}^{'2}$ are two chi-square variables with one degree of freedom. Then the pdf of
 $U \triangleq {\|\mathbf{e}_L' \|}_2^2$ is the convolution of these two chi-square pdfs, denoted as $f_U (u)$. The pdf of the localization error $e_L = {\|
\mathbf{e}_L' \|}_2 \triangleq V$ is given by $f_V (v) = 2 v f_U (v^2)$. As a result, the closed-form pdf of $e_{L}$, for given
sensor node positions, is given by
\begin{IEEEeqnarray}{rCl}
   p_{e_{L}}(x) &=& \frac{x}{\sigma_x \sigma_y} \exp \left(-\frac{x^2}{2
   \sigma_x^2}\right) \exp\left[- \frac{1}{2} \left(\frac{m_x^2}{\sigma_x^2} +
   \frac{m_y^2}{\sigma_y^2}\right)\right] \IEEEnonumber \\
   &&\times \sum_{i=0}^{\infty} \sum_{l=0}^{\infty} \left[ \frac{\Gamma(i+l+1/2)}{i! ~ l! ~ \Gamma(1/2+l)} \left(\frac{x m_y^2 \sigma_x^2}{2 m_x
   \sigma_y^4}\right)^{l} \right.
   \left.\left(\frac{x(\sigma_y^2 - \sigma_x^2)}{m_x \sigma_y^2}\right)^{i} \mathbf{I}_{i+l} \left(\frac{x
   m_x}{\sigma_x^2}\right)\right], x \geq 0
\label{pdf_2d}
\end{IEEEeqnarray}
where $m_x$, $m_y$ and $\sigma_x^2$, $\sigma_y^2$ are means and variances of $\hat{x}_p^{'}$ and $\hat{y}_p^{'}$, respectively, and $\Gamma(x)$ and $\mathbf{I}_{\alpha} (x)$ are Gamma function and $\alpha$th-order modified Bessel function of the first kind, respectively. The mean and variance of $e_{L}$, denoted as $m_{e_{L}}$ and $\sigma_{e_{L}}^2$, are calculated using their definitions, by numerical integration over the above pdf. Note that the above calculations are based on a particular fixed node placement. If the nodes are randomly placed within the area according to some distribution, say uniform distribution, integrating $2N$ times over all node coordinates gives the theoretical average performance.


\section{Distributed Implementation of WCL}
\label{sec:Distributed}
Based on the theoretical analysis, WCL requires large number of nodes in order to achieve a high level of accuracy. This causes large communication overhead in terms of number of messages and transmit power consumption when implemented in a centralized manner. We now present a practical distributed implementation of WCL to address this problem in the context of CR networks.

\subsection{Algorithm Description}
\label{subsec:Algorithm Description}
Clustering is commonly used in implementation of distributed algorithms. The challenge in designing these algorithms is to reduce the amount of information exchanged among clusters. The proposed algorithm accomplishes this by splitting the task into two phases. First, the cluster with the highest average RSS is selected. This cluster then acts as the coordinator for the second phase which performs the actual WCL algorithm.

\subsubsection{Clustering}
\label{subsubsec:Clustering}
Nodes, $N_j$, in the CR network are assumed to be organized into clusters, $C_i$, using the clustering algorithm presented in \cite{Zhang2002}. This particular algorithm is based on geographic boundaries and generates clusters which have a predefined shape (hexagonal in this case). The regularity in the shape of the cluster is beneficial for the first step of the distributed algorithm since WCL works best for a radially symmetric distribution of nodes. However, the proposed algorithm would work for any arbitrary clustering algorithm as long as nodes are clustered based on proximity. The PU is assumed to be randomly placed in the entire area and can thus be in any of these clusters.

Each cluster selects one node, $\head(C_i)$, to be the cluster head.  The cluster head maintains a table of all adjacent clusters, $\adj(C_i)$, and the location of its member nodes, $\mathbf{L}(N_j), N_j\in C_i$.

Once a PU is detected by a node, using an existing PU detection method such as energy detection \cite{Digham2007}, it transmits its RSS measurement to the corresponding cluster head which then participates in the distributed WCL algorithm outlined in the next section. Using a predetermined fusion rule, such as a majority vote, the cluster head makes a decision on the presence of the PU. All clusters that decide that the PU is present, participate in the Distributed WCL algorithm detailed in the next subsection. These clusters form a set defined as $C_{\textit{active}}$. If in addition, some prior knowledge of the PU transmit power is available, a situation typical of a CR network coexisting with a licensed incumbent network, then a predetermined threshold can be used to further reduce the number of clusters in $C_{\textit{active}}$. The threshold can be chosen based on the average node spacing to get a certain desired average number of nodes. We investigate the effect of this number on WCL accuracy in Sec.~\ref{subsec:VariableParticipation}.

\subsubsection{Distributed WCL (DWCL)}
\label{subsubsec:Algorithm}
The first phase, which selects the head cluster, is shown in Algorithm \ref{alg:selection}. The objective is to select the cluster with the highest average RSS, $\overline{\pw}(C_i)$. This is done in a distributed manner by comparing the average RSS between neighboring clusters. However, directly implementing this procedure would require a large amount of inter-cluster communication.

To solve this issue, a cluster only communicates its average RSS to the neighboring cluster, $\nex(C_i)$, in the direction of the PU relative to the cluster's centroid.

Although techniques for estimating the angle of arrival of a transmitter could be beneficial, these methods require cluster heads to have multiple antennas which increases complexity of sensors. In order to get an approximation of the PU direction, we propose a metric, $\hat{\mathbf{L}}(C_i)$, shown in Line 5 of Algorithm \ref{alg:selection}. We get $\hat{\mathbf{L}}(C_i)$ by subtracting the WCL result of all nodes within the same cluster from the geometric centroid of those nodes. This vector approximates the RSS gradient within the cluster. It corresponds to an approximate direction of the PU since on average the RSS in (\ref{received power}) decreases as distance to PU increases. Therefore, the negative of the gradient is directed towards the point with maximum RSS.

Once the cluster with the highest average RSS is selected, Algorithm \ref{alg:mwcl} is performed. In this step, the final WCL calculation is done using nodes within a given radius, $R^*$, of the selected cluster's strongest node (SN), $N_S$. $R^*$ is adaptively chosen to be the minimum of the distance to the closest map edge, $\edge(N_S)$, and the cluster radius $R_C$. This reduces the border effect, which is the tendency of the location estimate to be biased towards the center when the PU is close to the map border, by ensuring a radially symmetric node distribution even when the PU is close to the network edge.

In order to improve the localization accuracy, additional nodes from adjacent clusters are also included in the WCL calculation. Using the location of $N_S$ and $R^*$, these adjacent clusters can reduce the amount of information exchanged by only transmitting the location and RSS measurements of nodes that fall within this circle. In the case of large clusters, each cluster could also choose to calculate a weighted centroid using its own nodes and transmit the result of this calculation only. Since WCL is essentially an averaging algorithm, this technique allows even Algorithm \ref{alg:mwcl} to be distributed.
\begin{algorithm}
\caption{Head Cluster Selection}
\label{alg:selection}
\algsetup{indent=2em}
\begin{algorithmic}[1]
\FORALL{$C_i \in C_{\textit{active}}$}
    \STATE  $\overline{\pw}(C_i) \leftarrow \average(\pw(N_j))$
    \STATE  $\mathbf{L}_c(C_i) \leftarrow \average(\mathbf{L}(N_j))$ \COMMENT{cluster centroid}
    \STATE  $\mathbf{L}_w(C_i) \leftarrow \frac{\sum_{N_j \in C_i} \left[\pw(N_j) - \min(\pw(N_j))\right] \mathbf{L}(N_j)} {\sum_{N_j\in C_i}(\pw(N_j) - \min(\pw(N_j))}$ \COMMENT{cluster WCL}
    \STATE $\hat{\mathbf{L}}(C_i) \leftarrow \frac{\mathbf{L}_c(C_i) - \mathbf{L}_w(C_i)}{\|\mathbf{L}_c(C_i) - \mathbf{L}_w(C_i)\|}$ \COMMENT{direction of gradient}
    \STATE $\nex(C_i) \leftarrow \argmax_{C_j \in \adj(C_i)} \left(\frac{\mathbf{L}_c(C_j) - \mathbf{L}_c(C_i)}{\|\mathbf{L}_c(C_j) - \mathbf{L}_c(C_i)\|} \circ \hat{\mathbf{L}}(C_i)\right) $
    \IF{$\overline{\pw}(C_i) > \overline{\pw}(\nex(C_i))$}
        \IF{$\overline{\pw}(C_i) > \overline{\pw}(C_j)~\forall C_j \in \adj(C_i)$}
            \STATE Select $C_i \rightarrow C_{\textit{wcl}}$ \COMMENT{cluster for final WCL}
            \STATE Proceed to Algorithm \ref{alg:mwcl}
        \ENDIF
    \ENDIF
\ENDFOR
\end{algorithmic}
\end{algorithm}

\begin{algorithm}
\caption{Modified WCL}
\label{alg:mwcl}
\algsetup{indent=2em}
\begin{algorithmic}[1]
    \STATE $N_{S} \leftarrow \argmax_{N_i \in C_{\textit{wcl}}}{\pw(N_i)}$ \COMMENT{SN acts as center of WCL}
    \STATE $R^* \leftarrow \min(\edge(N_S),R_C)$ \COMMENT{border correction}
    \FORALL{$C_j \in \adj(C_{\textit{wcl}})$}
        \STATE $C_{\textit{wcl}}^* \Leftarrow \pollcluster(C_j,N_{S},R^*)$ \COMMENT{add nodes from adjacent clusters}
    \ENDFOR
    \STATE $\mathbf{L}_{est} \leftarrow \frac{\sum_{N_j \in C^*_{wcl}} \left[\pw(N_j) - \min(\pw(N_j))\right] \mathbf{L}(N_j)} {\sum_{N_j \in C^*_{wcl}}\left[\pw(N_j) - \min(\pw(N_j)\right]}$
\end{algorithmic}
\vspace{1em}
\begin{algorithmic}[1]
\newcommand{\algorithmicfunction}{\textbf{function}\ }
\newcommand{\algorithmicfunctionend}{\textbf{end function}\ }
    \STATE \algorithmicfunction $\pollcluster (C_i,N_{S},R^*)$
        \STATE \hspace{2em}\textbf{return} $C^*_i = \{N_j | N_j \in C_i$,~ $\left\|\mathbf{L}\left(N_{S}\right)-\mathbf{L}\left(N_{j}\right)\right\|\leq R^*\}$
    \STATE \algorithmicfunctionend
\end{algorithmic}
\end{algorithm}

The power efficiency and computational complexity of this distributed approach could further be improved by taking into account a practical mobility model of the PU. There has been a lot of work such as \cite{Hue2002}, which discussed how to incorporate past observations of the target position and a mobility model to improve localization. However, applying these methods introduces complexity to WCL and defeats the purpose of such a low complexity scheme.

Intuitively, assuming the PU continuously  transmits during successive WCL estimations, a PU would most likely be in the same cluster as it was in the previous WCL calculation, or it could move to an adjacent cluster. By considering this fact, Algorithm \ref{alg:selection} can further limit the clusters involved in choosing the head cluster, which is the most power consuming phase, by only taking the previous cluster head and its adjacent clusters as $C_{\textit{active}}$.

\subsection{Analysis of Communication Overhead}
\label{subsec:Communication Overhead}

We analyze the total transmit power and computational complexity of CWCL and the proposed DWCL algorithm. In order to do so, we first need to investigate the transmit power on a single link of length $d$. The minimum transmit power is obtained from (\ref{received power}) as $P_{t,\min}=P_{r,\min}d^{\gamma}10^{-s/10}$, where $P_{t,\min}$ is the minimum received power to correctly receive the message. For the computation complexity estimates, we assume the number of operations (OPS) for addition, substraction and comparison is 1; for multiplication and division is 10.

\subsubsection{Centralized WCL}

In CWCL, the total number of messages is the sum of reporting messages from nodes to the fusion center, thus, $N$. For the transmit power consumption, we assume the fusion center is roughly at the center of a circle area of radius $R$, and nodes are placed according to a uniform distribution. In this case, one can verify that, $d_i$, the distance from the $i$th node to the fusion center satisfies $d_i\sim \sqrt{\mathcal{U} (0, R^2)}$, where $\mathcal{U}(a,b)$ is the uniform distribution from $a$ to $b$. Thus the average total power consumption is expressed as
\begin{equation}\label{total_power_central}
   \E [ P_{t,c} ] = \E [ P_{r,\min} \sum_{i=1}^{N} (d_i^{\gamma}10^{-s_i/10}) ] \approx N P_{r,\min} \E [d_i^{\gamma}] \E [10^{-s_i/10}],
\end{equation}
where we use statistical average to replace sum of realizations.

Note that $\E [d_i^{\gamma}] = \E [{(d_i^2)}^{\gamma/2}]$, which is the $\gamma/2$th moment of an uniform variable. In order to calculate $\E [10^{-s_i/10}]$, we need the distribution of $10^{-s_i/10}$, which is a function of central Gaussian variable $s_i$. Thus we can derive the desired pdf as \cite{Proakis2001}
\begin{equation}\label{gaussian_derive}
   p (x) = \frac{10}{\ln{10} \sqrt{2 \pi} \sigma_s x} \exp{\left(-50 \frac{\log{y}^2}{\sigma_s^2}\right)}.
\end{equation}

Finally, the average total power consumption is given by
\begin{equation}\label{total_power_central2}
   \E [ P_{t,c} ] = N P_{r,\min} \E [d_i^{\gamma}] \int_0^{\infty} x p(x) d x.
\end{equation}

From the definition of WCL estimate, number of OPS of CWCL is $\mathcal{O}(25N)$.

\subsubsection{Distributed Algorithm}

Assume that: $M$ is the average number of nodes in one cluster; $L=N/M$ is the average number of clusters; $K$ is the number of neighboring clusters for each cluster; $\eta$ is the average percentage of clusters who decide they are the head cluster in step 7 of Algorithm 1 (clusters that have to perform step 8).

The total number of messages for Algorithm 1 is $\mathcal{O}(ML+2L+2\eta KL)$. To calculate transmit power, in step 1) we use the same procedure as CWCL (with a smaller radius); step 2) and 3) are communication among cluster heads, we use the average distance between centers of two clusters as the estimate of the average distance among cluster heads. In Algorithm 2, the total number of messages is $\mathcal{O}(2K)$ and the calculation of power consumption for this process is the same as step 2) of Algorithm 1.

In terms of computational complexity, the number of OPS for Algorithm 1 is $\mathcal{O}(27N+44L+64KL+\eta KL)$ while the number of OPS for Algorithm 2 is $\mathcal{O}(34KM+26M)$.


\section{Evaluation of WCL Performance}
\label{sec:Evaluation}

In this section, we evaluate the performance of CWCL algorithm and its proposed distributed implementation. For CWCL algorithm, using the theoretical framework presented in Section \ref{sec:Analysis}, we consider impact of sensor node positioning error, different levels of shadowing $\sigma_{s}$ and different correlation distances $X_C$. The effect of node placement is investigated under random grid and uniform distribution.

\subsection{Performance Metric}
\label{subsec:PerformanceMetric}
Prior work on WCL uses mean localization error as the performance metric of interest. In \cite{Liu2009}, a slightly modified performance metric where the mean error is normalized by the transmission range $R$ of PU is adopted. Simulations in these prior work have shown that the accuracy of WCL improves as the density of nodes is increased. This result is intuitive since the node spacing decreases as the density of nodes is increased. In this work, we introduce a novel performance metric where the mean localization error $m_{e_L}$ is normalized with respect to the average node spacing $D$. The motivation for choosing this metric is to eliminate the effect of geometry scaling which intuitively improves accuracy.

\subsection{Performance of CWCL Algorithms}
\subsubsection{Impact of Shadowing}
\label{subsec:Shadowing}

\begin{figure}
\centering
\subfloat[]{
\includegraphics[width=0.48\columnwidth]{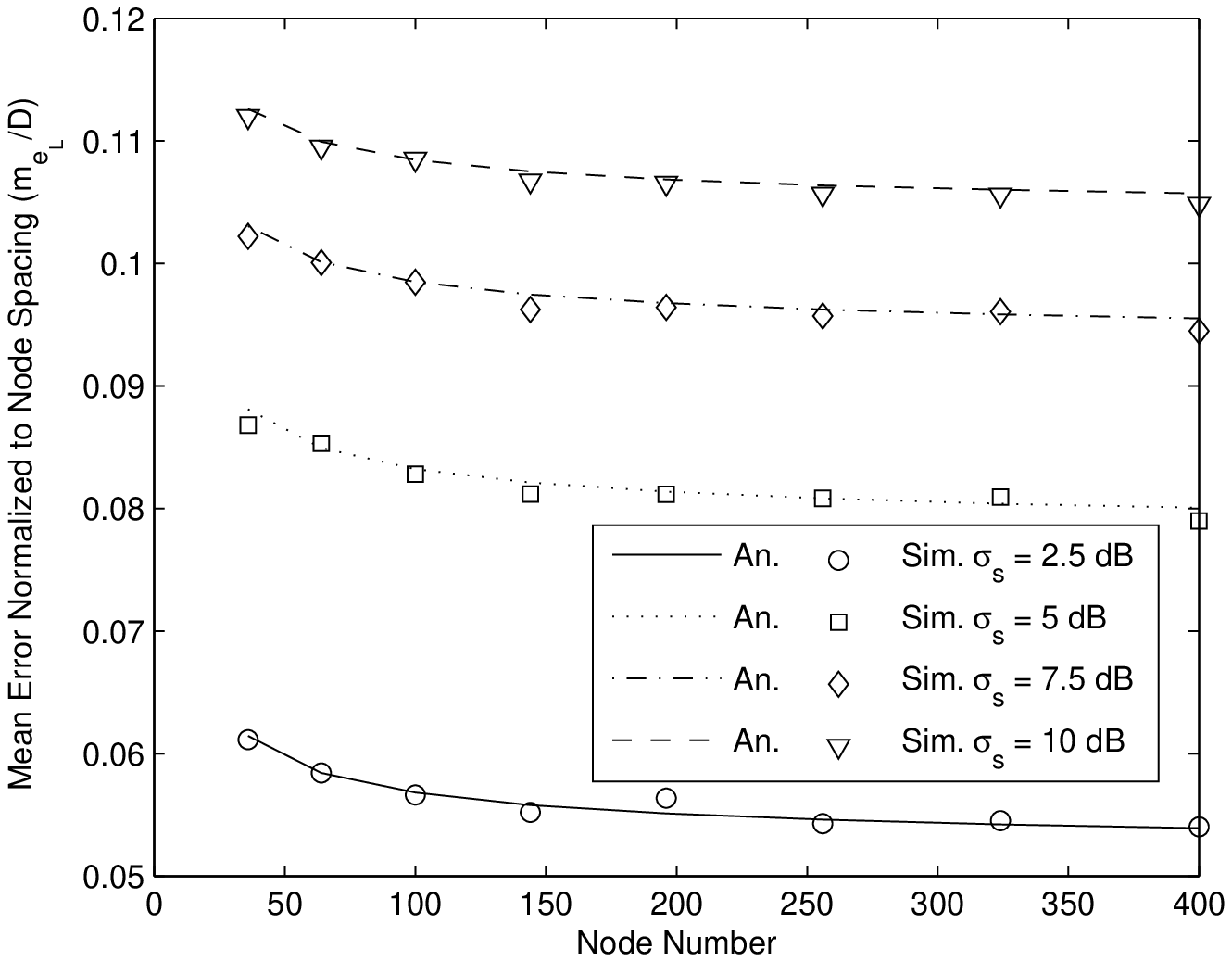}
\label{fig:FixGridSquare_Shadow_Mean}
}
\subfloat[]{
\includegraphics[width=0.48\columnwidth]{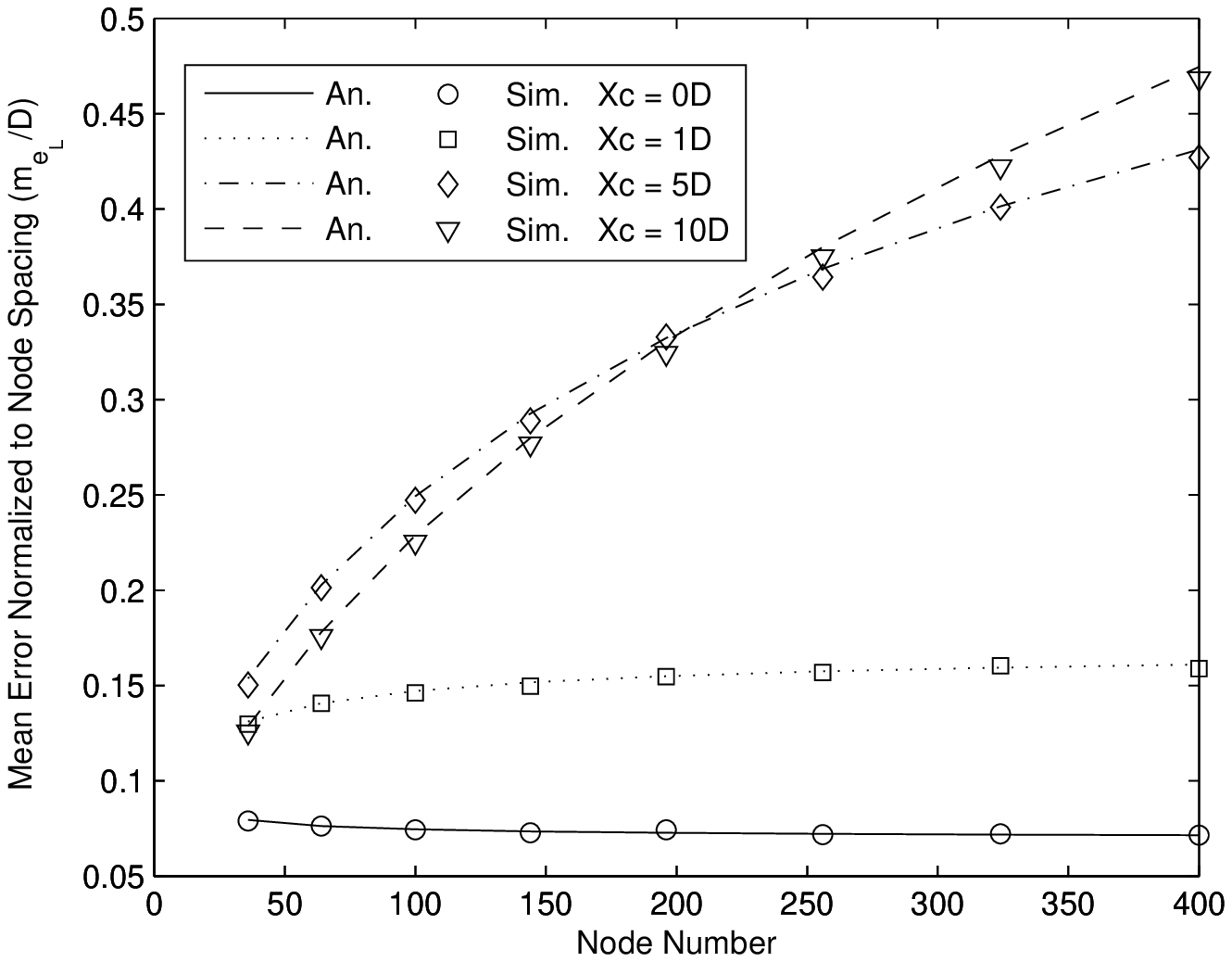}
\label{fig:FixGridSquare_Corr_Mean}
}
\caption{WCL analysis and simulation results for the normalized mean localization error in  \subref{fig:FixGridSquare_Shadow_Mean} uncorrelated and \subref{fig:FixGridSquare_Corr_Mean} correlated shadowing environment with varying levels of shadowing, $\sigma_s$. For this scenario, nodes have no positioning error ($\sigma_l=0 \text{m}$), and are placed on a grid with equal spacing (fixed grid placement) in an area with $R=100$m. PU is at the center of the area.}
\label{fig:FixedGrid}
\end{figure}

In order to isolate the effect of shadowing on WCL performance we consider a fixed grid placement of nodes with perfect positioning (i.e. $\sigma_l=0$) and fixed PU position.  The normalized mean localization error for this scenario is shown in Fig.~\ref{fig:FixGridSquare_Shadow_Mean}. For given shadowing variance, as the number of nodes increases, there is a slight improvement in the accuracy, which is a consequence of averaged shadowing over many nodes. Note that this gain saturates at approximately 200 nodes or more. We also observe that WCL is quite robust to shadowing, since for an additional $7.5\,\text{dB}$ of shadowing from $2.5\,\text{dB}$ to $10\,\text{dB}$, normalized error increases only 5\%. Therefore WCL is well suited for PU localization in severely shadowed environments.

\subsubsection{Impact of Correlation}
\label{subsec:Correlation}

The results for mean localization error under correlated shadowing are shown in Fig.~\ref{fig:FixGridSquare_Corr_Mean}. assuming fixed value of $\sigma_s = 4\,\text{dB}$, and different levels of $X_C$ normalized to $D$ are presented. In highly correlated shadowing such as $X_C > 5D$, increasing the number of nodes always worsens the error. For the high node density case, e.g. 400 nodes, the average error is as high as 47\% of $D$. We conclude that when all nodes are included in the WCL calculation, correlation has a degrading effect to accuracy. Further investigation shows that the correlation distance resulting in the worst performance is proportional to the number of nodes involved.

\subsubsection{Impact of Node Positioning Error}

\begin{figure}
\centering
\includegraphics[width=0.6\columnwidth]{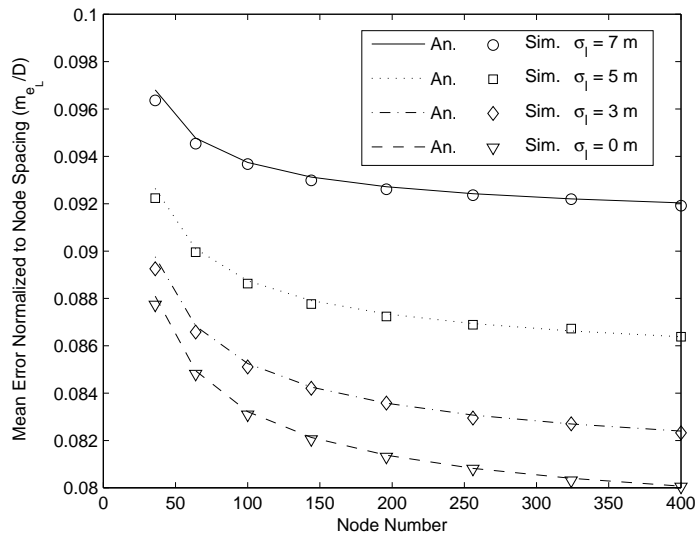}
\caption{WCL analysis and simulation results for the normalized mean localization error in an uncorrelated shadowing environment ($\sigma_s = 5 \text{dB}$) as a function of variance in the participating nodes' positioning error, $\sigma_l$. For this scenario, nodes are placed on a grid with equal spacing (fixed grid placement) in an area with $R=100$m. PU is at the center of the area.}
\label{fig:Imperfect Location Knowledge}
\end{figure}

The impact of positioning accuracy of sensing nodes is presented in Fig.~\ref{fig:Imperfect Location Knowledge} with a fixed grid placement of nodes in an i.i.d. shadowing environment with $\sigma_s=5 \text{dB}$. Analysis shows that errors in sensor node positioning decrease the WCL accuracy but the impact is marginal. For example, $\sigma_l$ of $7$m increases the error for about $1$\%.

\subsubsection{Impact of Node Placement}
\label{subsec:NodePlacement}

The previous subsections considered fixed node placement and PU location. Realistically, node placement can be random and the PU also appears at random locations relative to the sensor nodes. Next we consider two cases of random node placement: random grid and uniformly distributed placement.

\begin{figure}
\centering
\includegraphics[width=0.6\columnwidth]{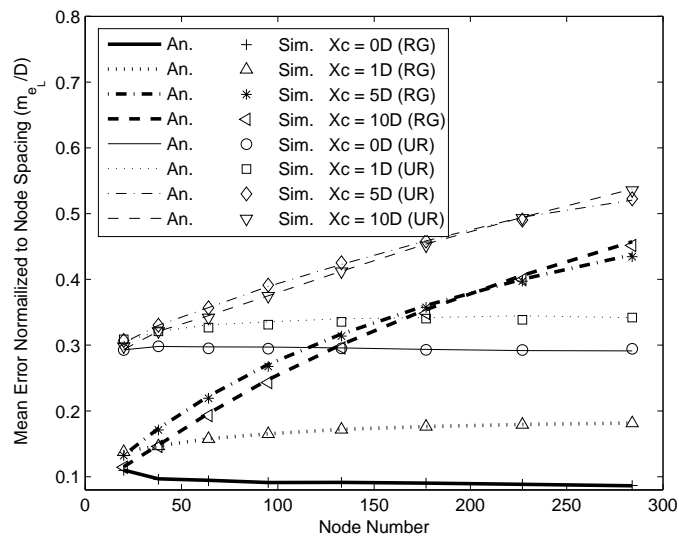}
\caption{WCL analysis and simulation results for the normalized mean localization error in a correlated shadowing environment ($\sigma_s=4 \text{dB}$) with varying correlation distance normalized to the node spacing. Results of two placements, random grid (RG) and  uniform random (UR), are shown in an area with $R=100$m. For both placements, $\sigma_l=0 \text{m}$.}
\label{fig:RandomGridandUniformRandomPlacement}
\end{figure}

In the random grid scenario, the sensor nodes are still on a grid, but the PU can appear uniformly within the center box. A grid distribution of nodes can be practically implemented if nodes are explicitly deployed for the purpose of acting as a dedicated localization sensor network. Evenly placed beacons have been shown to provide better accuracy for WCL in WSNs \cite{Salomon2005}. Results for random grid placement in Fig.~\ref{fig:RandomGridandUniformRandomPlacement} show approximately 10\% performance loss compared to fixed PU location scenarios.

The uniformly distributed placement has been used in \cite{Laurendeau2010,Blumenthal2007,Liu2009} to more accurately model scenarios where the sensor node positions are not predefined, such as in mobile ad-hoc networks. The results for uniformly distributed placement are also shown in Fig.~\ref{fig:RandomGridandUniformRandomPlacement}. It is observed that randomness in the node placement can increase the error as much as three times compared to the fixed node placement.

\subsubsection{Impact of Degree-of-Irregularity}
\label{subsec:VariableDOI}

The results given so far assumed an ideal circular coverage area for the PU. Works such as \cite{Bulusu2000,He2003} indicated that in practical scenarios, the circular model cannot be guarranteed due to several non-idealities such as fading, shadowing, and interference in the system. Intuitively, the Degree-of-Irregularity (\emph{DOI}) parameter determines how closely the coverage area can be approximated by a circle. \emph{DOI} = 0 means the transmission range is perfectly circular. The model used in our simulation is the \emph{DOI} model briefly described in \cite{He2003} where the transmission range is a correlated Gaussian r.v. centered at $R=100\,\text{m}$ with variance $\sigma=R(\textit{DOI})$.

\begin{figure}
\centering
\includegraphics[width=0.6\columnwidth]{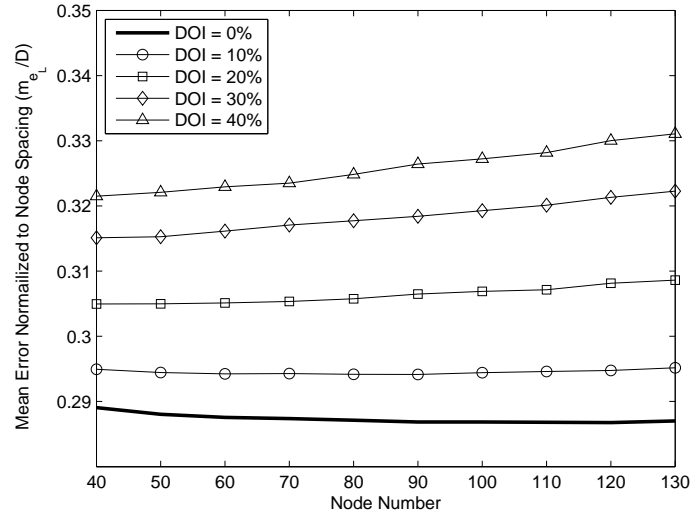}
\caption{Simulation results for the normalized mean error in an uncorrelated shadowing environment with $\sigma_s=4\,\text{dB}$ with varying Degree-of-Irregularity (DOI). For this scenario, nodes are uniformly and independently distributed (uniform random placement) in an area with average $R=100\,\text{m}$.}
\label{fig:VaryDOI}
\end{figure}

In Fig.~\ref{fig:VaryDOI} the effect of \emph{DOI} is investigated in the presence of uncorrelated medium level shadowing ($\sigma_s=4\,\text{dB}$). The result shows that \emph{DOI} between 0\% and 40\% increases the error by 4\%. \emph{DOI} is not a significant impairment for WCL because nodes closest to the PU, where the effect of irregularity is negligible, are weighted more than nodes closer to the edge of the coverage area. However, a slight increase in the normalized mean error can be observed as more nodes are used at high DOI (30\% or 40\%). The denser the node distribution, the more nodes are included in WCL, but the more nodes there are, the more deviation from a perfect circular coverage coverage area, leading to lower accuracy.

%


\subsubsection{Variable Participation}
\label{subsec:VariableParticipation}
From Fig.~\ref{fig:RandomGridandUniformRandomPlacement} we observe that in correlated shadowing increasing the number of nodes also increases the error. One solution to this problem is to reduce the number of participating nodes used in WCL for this scenario. This could be achieved by ignoring nodes farther away from the PU to reduce the number of nodes resulting in less degradation.

\begin{figure}
\centering{
\includegraphics[width=0.6\columnwidth]{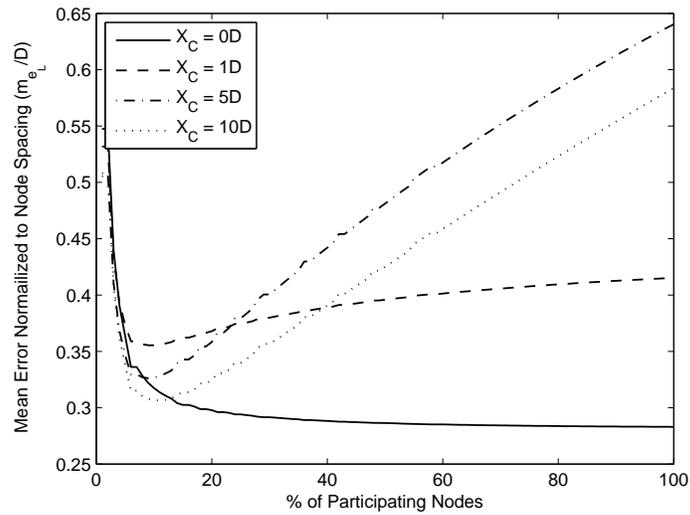}}
\caption{Effect of varying the number of participating nodes to WCL accuracy, with different correlation distances ($X_C$) . In this scenario, $N=100$ uniformly and independently distributed nodes (uniform random placement) are in range of the PU, while the percentage of these $N$ nodes that are included in the WCL calculation is varied ($R=100\,\text{m}, \sigma_s=4\,\text{dB}, \sigma_l=0\,\text{m}$).}
\label{fig:VaryingParticipation}
\end{figure}

In Fig.~\ref{fig:VaryingParticipation} we investigate varying levels of node participation for a fixed average node spacing, $D$. In this case the total number of nodes is 100, and the density remains the same while the number of involving nodes is varied. To select which nodes are included in the WCL calculation, nodes are sorted according to their RSS values. The nodes with high received power are included in WCL. As a result, increasing the participation for the uncorrelated case always improves the accuracy. Thus, 100\% participation is optimal. However, even with a little correlation, this number is greatly reduced and about 10\% of the total number of nodes is optimal. Another selection scheme was previously proposed in \cite{Chandra2007} where only nodes whose RSS is within 15\% of the stronges node's RSS are used.

\subsubsection{Comparison With Other Localization Techniques}

The localization accuracy of WCL is also compared to other techniques suitable for CR scenarios. This includes lateration, centroid and strongest node (SN). Lateration is a range-based technique that uses least-squares estimation using distance estimates to the target. The centroid algorithm, on the other hand, is a range-free technique which calculates the average of all node positions in range of the PU. Finally, the SN technique simply selects the node with the highest RSS and uses its location as an estimate of the PU position.

\begin{figure}
\centering
\includegraphics[width=0.6\columnwidth]{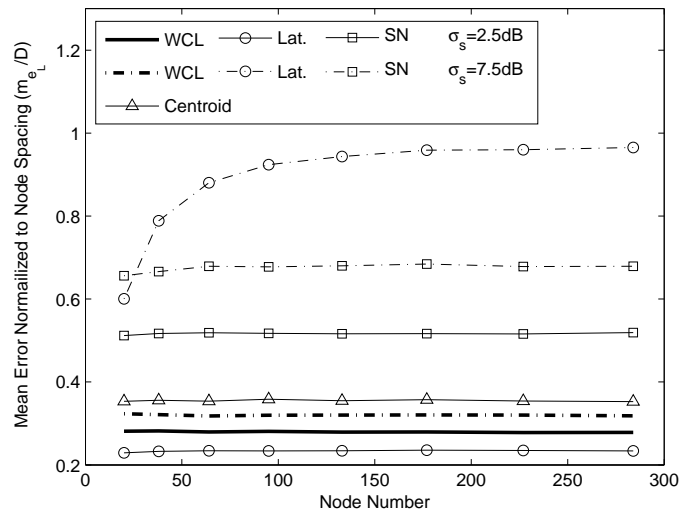}
\caption{Comparison of WCL performance with a range-based scheme (Lateration) and a range-free scheme (Centroid) under different shadowing variance, $\sigma_{s}$. For this scenario, nodes are uniformly and independently distributed (uniform random placement) in an area with $R=100\,\text{m}$ (uncorrelated shadowing). Note that the Centroid technique is unaffected by $\sigma_s$.}
\label{fig:Comaprison With Other Techiques}
\end{figure}

Results in Fig.~\ref{fig:Comaprison With Other Techiques} show that WCL performs better under higher $\sigma_{s}$ compared to lateration. Although this seems counter-intuitive at first, the result can be explained by taking note that range-based techniques require an accurate estimate of the distance to the target to provide accurate results. In the case of non-interactive localization, where the only available information is the RSS from the PU, shadowing induces a very large error in the range estimates. Under low shadowing ($<2.5$ dB), the lateration method is more accurate than WCL. At 0 dB shadowing (not shown in the figure), the range-based technique will achieve perfect localization. This result supports the claim that WCL is a viable method in the context of non-interactive localization because of its robustness. The centroid algorithm performs slightly worse than WCL because it does not utilize RSS information. However, the performance difference becomes smaller at higher level of shadowing.

\subsection{Performance of DWCL algorithms}

\subsubsection{Comparison of Accuracy}
\label{subsubsec:Accuracy}
\begin{figure}
\centering{
\includegraphics[width=0.6\columnwidth]{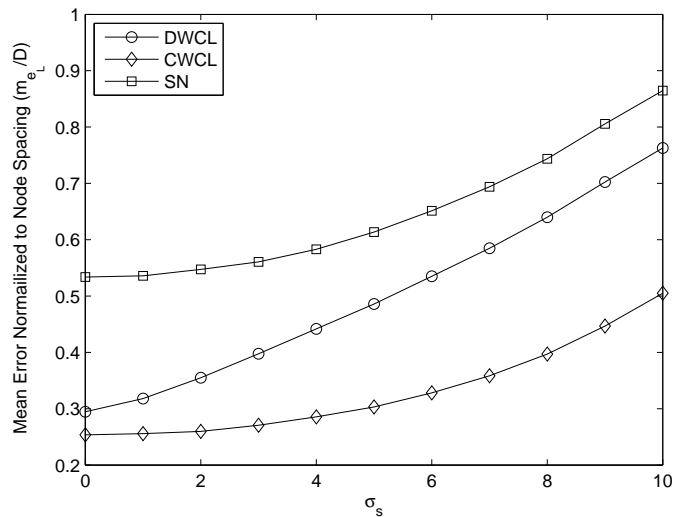}}
\caption{Performance comparison of centralized/distributed WCL (CWCL/DWCL) and strongest node (SN) algorithms with varying $\sigma_s$. In this scenario, $N=1000$ uniformly and independently distributed nodes (uniform random placement) in a $2000\,\text{m} \times 2000\,\text{m}$ square area. An average cluster radius of $R_C=200\,\text{m}$ and  the PU is placed randomly in the entire area with uniform probability ($X_C=0\,\text{m}, \sigma_l=0\,\text{m}$).}
\label{fig:dwcl compare}
\end{figure}

Fig.~\ref{fig:dwcl compare} presents simulation results that compare normalized mean error of CWCL, SN and DWCL under different shadowing environments. SN is included in the comparison because it forms the basis of the head cluster selection for DWCL so it acts as an upperbound to the error. We observe that the proposed DWCL performance is between CWCL and SN algorithms. It is also interesting to note that the accuracy of DWCL degrades relative to CWCL and approaches the performance of SN when shadowing is high. The reason behind this is that for cluster head selection, DWCL is dependent on the SN algorithm. The accuracy of the SN degrades at large shadowing shown in Fig.~\ref{fig:Comaprison With Other Techiques}.

\subsubsection{Comparison of Transmit Power}
\label{subsubsec:Power}

\begin{figure}
\centering{
\includegraphics[width=0.6\columnwidth]{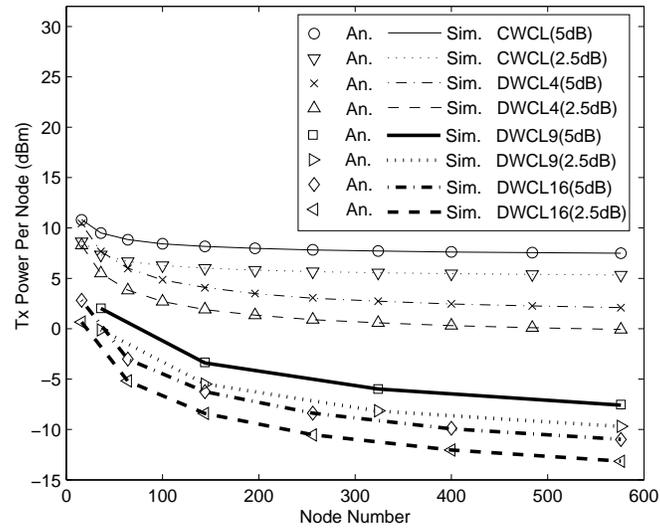}}
\caption {Analysis and simulation results for average transmit power per node for CWCL and DWCL algorithms with varying levels of shadowing with fixed grid placement ($R=100\,\text{m}, \sigma_l=0\,\text{m}$).}
\label{fig:FixGrid_NodePower_Theo}
\end{figure}

Fig.~\ref{fig:FixGrid_NodePower_Theo} compares the transmit power consumption for CWCL and DWCL with different number of clusters. In the simulation we set $P_{r,min} = -70 \text{dBm}$ which is the mean received power at the border of the area of $R=100\,\text{m}$, assuming $P_0=0$ dBm, $d_0=1$ m and $\gamma=3.8$ in (\ref{received power}). We observe that DWCL with 16 clusters consumes about $15$dBm less power for each node compared with CWCL. The analysis also shows that transmit power of each node is reduced as the number of node increases, since the average communication distance is getting smaller in DWCL.

\subsubsection{Comparison of Number of OPS}
\label{subsubsec:OPS}

\begin{figure}
\centering{
\includegraphics[width=0.6\columnwidth]{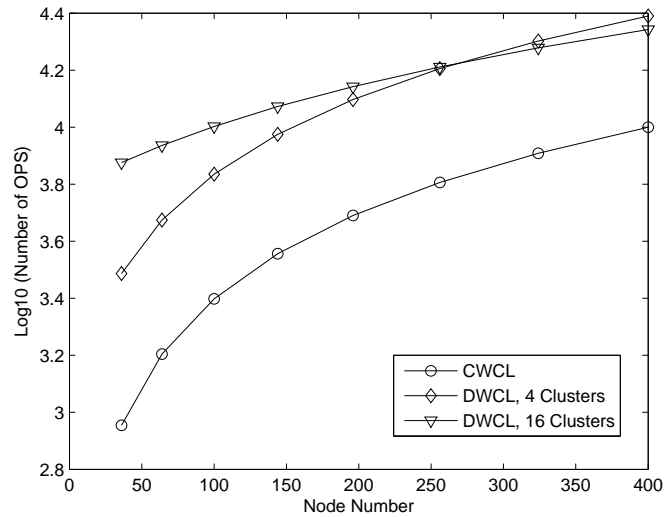}}
\caption {Number of operations (OPS) for CWCL and DWCL for different cluster sizes.}
\label{fig:FixGrid_CPU_Theo}
\end{figure}

Fig.~\ref{fig:FixGrid_CPU_Theo} compares computational complexity estimated in terms of number of OPS. It shows that distributed algorithms require about three times more operations than CWCL, however total transmit power is reduced.
%


\section{Conclusion}
\label{sec:Conclusion}
In this paper we present the first analytical framework for the performance of WCL for localization of PU transmitters in CR networks.  We derived the closed-form expression for the error distribution of WCL parameterized by node density, node placement, shadowing variance, correlation distance and errors in sensor nodes positions.  Using the theoretical results in conjunction with numerical simulations, the robustness of WCL against various physical conditions is investigated and quantified. Based on the above analysis, we provide guidelines in terms of number of nodes and node placement required to achieve certain localization accuracy using WCL and propose variable node participation for improved WCL performance. To facilitate the deployment of WCL in an energy efficient context, a distributed cluster-based implementation of WCL is also proposed, which does not assume a fusion center that collects all RSS information from nodes. The distributed method achieves comparable accuracy with its centralized counterpart, and greatly reduces total power consumption.


\appendix[Correlation Coefficient Between Localization Errors In X And Y Dimensions]
\label{app:CorrelationCoefficient}

We derive the correlation coefficient between localization errors in x and y dimensions, $\rho_{\hat{x}_p \hat{y}_p}$, which is given by
\begin{equation}
   \rho_{\hat{x}_p \hat{y}_p} \sigma_{\hat{x}_p} \sigma_{\hat{y}_p}
            = \E [\hat{x}_p \hat{y}_p]-m_{\hat{x}_p}m_{\hat{y}_p}.
\label{dimensional correlation}
\end{equation}

For notational convenience we define $q_i \triangleq P_i - P_{\min} \sim \mathcal{N}(\mu_i,\sigma_s^2)$ and form the vector $\mathbf{q} \triangleq \left[q_1, q_2, \ldots, q_{M}\right]^{T}$, so that $\mathbf{q} \sim \mathcal{N} (\boldsymbol{\mu},\mathbf{\Omega})$ with $\mathbf{\Omega}_{ij} \triangleq \sigma_s^2 e^{- \| \overline{\mathbf{L}}_i-\overline{\mathbf{L}}_j \|/X_c}$. Note that this definition includes both i.i.d and correlated shadowing case. The first term in (\ref{dimensional correlation}) then becomes
\begin{equation}\label{reform}
   \E [\hat{x}_p \hat{y}_p] = \E \left[\frac{\mathbf{x}^{T} \mathbf{q} \mathbf{q}^{T} \mathbf{y}}{\mathbf{1}^{T} \mathbf{q} \mathbf{q}^{T}
   \mathbf{1}}\right].
\end{equation}

Define three new variables $t_1=\mathbf{x}^{T} \mathbf{q}$, $t_2=\mathbf{y}^{T} \mathbf{q}$ and $t_3=\mathbf{1}^{T} \mathbf{q}$, and their vector form $\textbf{t}=[t_1, t_2, t_3]^{T}$. Now the first term in (\ref{dimensional correlation}) becomes $\E [(t_1 t_2)/t_3^2]$. As shown in Sec. II,  $\mathbf{x} \sim \mathcal{N}(\overline{\mathbf{x}}, \sigma_l^2 \textbf{I})$, and $\mathbf{y} \sim \mathcal{N}(\overline{\mathbf{y}}, \sigma_l^2 \textbf{I})$. $t_1$ can be approximated as a single Gaussian variable as shown in Sec. \ref{subsubsec:IIDShadowing} since it is the same as $a$ in (\ref{est x}). We denote its distribution as $t_1 \sim \mathcal{N}(m_1, \sigma_1^2)$, similarly $t_2 \sim \mathcal{N}(m_2, \sigma_2^2)$. The distribution of $t_3$ is given as $t_3 \sim \mathcal{N}(m_3, \sigma_3^2)$, where $m_3=\mathbf{1}^{T} \boldsymbol{\mu}$ and $\sigma_3^2 = \mathbf{1}^{T} \mathbf{\Omega} \mathbf{1}$.

As a result $\textbf{t} \sim \mathcal{N} (\overline{\textbf{t}}, \mathbf{\Omega}_{\textbf{t}})$, where $\overline{\textbf{t}} = [m_1, m_2, m_3]^T$ is the mean vector and $\mathbf{\Omega}_{\textbf{t}}$ is the covariance matrix. To derive $\mathbf{\Omega}_{\textbf{t}}$ we find $\text{Cov} (t_1, t_2)$ as
\begin{equation}
\label{Cov}
    \text{Cov} (t_1, t_2) = \E [t_1 t_2] - m_1 m_2
                         = \E [(\sum_{i=1}^{N} x_i q_i)(\sum_{j=1}^{N} y_j q_j)] - m_1 m_2
= \overline{\textbf{x}}^{T} \mathbf{\Omega} \overline{\textbf{y}} - m_1 m_2
\end{equation}
where in the last the equation we used the fact $\E [x_i q_i y_j q_j] = \overline{x}_i \overline{y}_j \mathbf{\Omega}_{ij}$. Similarly we calculate all entries of $\mathbf{\Omega}_{\textbf{t}}$.

In order to simplify the calculation of (\ref{reform}), we perform the following transformation to $\textbf{t}$, $\textbf{t} = \overline{\textbf{t}} + \mathbf{\Omega}_{\textbf{t}}^{-1/2} \widetilde{\textbf{t}}$, thus $\widetilde{\textbf{t}} \sim \mathcal{N}(\mathbf{0}, \textbf{I})$. Clearly $t_i = m_i + \boldsymbol{\omega}_{i}^{T} \widetilde{\textbf{t}}, i=1,2,3$, where $\boldsymbol{\omega}_{i}$ is the $i$th column of $\mathbf{\Omega}_{\textbf{t}}^{-T/2}$. Then we have
\begin{equation}\label{results}
   \E [\frac{t_1 t_2} {t_3^2}] = \E [\frac{(m_1 + \boldsymbol{\omega}_{1}^{T} \widetilde{\textbf{t}})(m_2 + \boldsymbol{\omega}_{2}^{T} \widetilde{\textbf{t}})}{(m_3 + \boldsymbol{\omega}_{3}^{T} \widetilde{\textbf{t}})^2}].
\end{equation}

Since the expectation in (\ref{results}) involves multiplication and division of three correlated random variables, the calculation can not be done directly. In order to reduce the number of variables involved, first we form $\mathbf{A} =
[\boldsymbol{\omega}_{3}, \boldsymbol{\omega}_{1}, \boldsymbol{\omega}_{2}]$, then perform QR factorization to reduce dimensionality, obtaining $\mathbf{Q}^{T} \mathbf{A}  = \mathbf{R}$, in which $\mathbf{Q}$ is orthogonal and $\mathbf{R}$ is upper triangular
\begin{equation}\label{upper-tri_new}
   \mathbf{R} =               \left[\begin{array}{ccc}
                               r_{11} & r_{12} & r_{13} \\
                               0      & r_{22} & r_{23} \\
                               0      & 0      & r_{33}
                              \end{array}\right].
\end{equation}

From the orthogonal-invariant property of standard Gaussian variables \cite{Proakis2001}, the distribution of $\mathbf{Q}^{T} \widetilde{\textbf{t}}$ is the same as that of $\widetilde{\textbf{t}}$, which is denoted as $\mathbf{Q}^{T} \widetilde{\textbf{t}} \doteq \widetilde{\textbf{t}}$.

Note that since $\boldsymbol{\omega}_{3}^{T} \mathbf{Q} \mathbf{Q}^{T} \widetilde{\textbf{t}} = \boldsymbol{\omega}_{3}^{T} \widetilde{\textbf{t}}$, denominator of (\ref{results}) can
further be expressed as
\begin{equation}
    (m_3 + \boldsymbol{\omega}_{3}^{T} \mathbf{Q} \mathbf{Q}^{T} \widetilde{\textbf{t}})^{2}
    \doteq \left(m_3 + \boldsymbol{\omega}_{3}^{T} \mathbf{Q} \widetilde{\textbf{t}}\right)^{2}
    \doteq \left(m_3 + \left[r_{11}, 0, 0\right] \widetilde{\textbf{t}}\right)^{2}
    \doteq \left(m_3 + r_{11} \widetilde{t}_{1}\right)^{2},
\label{den2_new}
\end{equation}
where we use the fact that $\mathbf{Q}^{T}
\boldsymbol{\omega}_{3}$ is the first column of $\mathbf{R}$; and $\widetilde{\textbf{t}} \triangleq [\widetilde{t}_{1}, \widetilde{t}_{2}, \widetilde{t}_{3}]^{T}$.
Similarly, numerator of (\ref{results}) can also be further expressed as
\begin{IEEEeqnarray}{rCl}
(m_1 + \boldsymbol{\omega}_{1}^{T} \widetilde{\textbf{t}})(m_2 + \boldsymbol{\omega}_{2}^{T} \widetilde{\textbf{t}})
   &\doteq& \left(m_1 + r_{12} \widetilde{t}_{1} + r_{22} \widetilde{t}_{2}\right) \left(m_2  + r_{13} \widetilde{t}_{1} + r_{23} \widetilde{t}_{2} + r_{33} \widetilde{t}_{3}\right)
\label{num2_new}
\end{IEEEeqnarray}

Therefore, the expectation in (\ref{reform}) becomes
\begin{equation}
   \E [\hat{x}_p \hat{y}_p]
   = \frac {r_{12} r_{13}} {r_{11}^2} \E \left[\frac{\widetilde{t}_{1}^2+ \left(\frac{m_1}{r_{12}} + \frac{m_2}{r_{13}}\right) \widetilde{t}_{1} + \left(\frac{m_1 m_2 + r_{22} r_{23}}{r_{12} r_{13}}\right)}{\left(\widetilde{t}_{1} + \frac{m_3}{r_{11}}\right)^{2}}\right].
\label{exp1_new}
\end{equation}
After factorization (\ref{exp1_new}) becomes
\begin{equation}\label{corr-coeff_new}
\!\!\! \E [\hat{x}_p \hat{y}_p] = \frac {r_{12} r_{13}} {r_{11}^2} \!\!\left(\!\!1+ \frac{\left(\frac{m_1}{r_{12}} + \frac{m_2}{r_{13}}\right)-2 \frac{m_3}{r_{11}}}{\frac{m_3}{r_{11}}} + \frac{\left(\frac{m_1 m_2 + r_{22} r_{23}}{r_{12} r_{13}}\right)+\left({\frac{m_3}{r_{11}}}\right)^2-\left(\frac{m_1}{r_{12}} + \frac{m_2}{r_{13}}\right) \frac{m_3}{r_{11}}}{1+\left({\frac{m_3}{r_{11}}}\right)^2} \right).
\end{equation}
Substituting (\ref{corr-coeff_new}) into (\ref{dimensional correlation}) gives the result.


\bibliographystyle{IEEEtran}

\end{document}